\documentclass[%
 reprint,
 showpacs,
 §preprintnumbers,
 amsmath,amssymb,
 aps,
]{revtex4-1}

\usepackage{graphicx}
\usepackage{dcolumn}
\usepackage{bm}
\usepackage{multirow} 

\begin{document}

\preprint{APS/123-QED}


\title{Linear and nonlinear market correlations: \\characterizing financial crises and portfolio optimization}

\author{Alexander Haluszczynski}
\email{alexander.haluszczynski@gmail.com}
\affiliation{Ludwig-Maximilians-Universit\"at, Department of Physics, Schellingstra{\ss}e 4, 80799 Munich\\
 risklab GmbH, Seidlstra{\ss}e 24, 80335, Munich
}%

\author{Ingo Laut, Heike Modest, Christoph R\"ath}
\affiliation{
 Deutsches Zentrum f\"ur Luft- und Raumfahrt, Institut f\"ur Materialphysik im Weltraum, M\"unchner Str. 20, 82234 We{\ss}ling}%

\date{\today}

\begin{abstract}
Pearson correlation and mutual information based complex networks of the day-to-day returns of US S\&P500 stocks between 1985 and 2015 have been constructed in order to investigate the mutual dependencies of the stocks and their nature. We show that both networks detect qualitative differences especially during (recent) turbulent market periods thus indicating strongly fluctuating interconnections between the stocks of different companies in changing economic environments. A measure for the strength of nonlinear dependencies is derived using surrogate data and leads to interesting observations during periods of financial market crises. In contrast to the expectation that dependencies reduce mainly to linear correlations during crises we show that (at least in the 2008 crisis) nonlinear effects are significantly increasing. It turns out that the concept of centrality within a network could potentially be used as some kind of an early warning indicator for abnormal market behavior as we demonstrate with the example of the 2008 subprime mortgage crisis. Finally, we apply a Markowitz mean variance portfolio optimization and integrate the measure of nonlinear dependencies to scale the investment exposure. This leads to significant outperformance as compared to a fully invested portfolio.
\end{abstract}

\pacs{05.45.Tp, 89.65.Gh, 89.75.Hc}

\maketitle


\section{\label{sec:level1}Introduction\protect}

Investigating phenomena in the financial markets has been becoming increasingly popular in the physics community. Econophysicists unfold a new perspective \cite{mantegna1999introduction} complimentary to traditional approaches in finance and financial mathematics through leveraging the powerful tools from statistical physics such as random matrix theory \cite{laloux1999noise} or agent based market models \cite{giardina2003bubbles}. 

It is vital for various applications in finance to gain a comprehensive understanding of how financial assets move together, e.g. when assessing the risk associated with a portfolio. In order to do so, it is common practice to express mutual dependencies of financial assets in terms of the Pearson correlation coefficient of their return time series. 

Mantegna and Stanley \cite{mantegna1995scaling} showed the power law scaling behavior of the probability distribution of financial indices. Hsieh \cite{hsieh1995nonlinear} pointed out that returns of financial assets are not autocorrelated while their absolute values strongly are. Further studies pointed out the intermittent behavior of financial time series and how they resemble phenomena that we know from turbulence \cite{ghashghaie1996turbulent,Mantegna:1996aa}. These results show the nonlinear nature of financial time series and thus strongly indicate that linear measures for correlations might not be sufficient to fully describe the data.

Mantegna \cite{mantegna1999hierarchical} first proposed the concept of Minimum Spanning Trees (MST) based on linear correlations between stocks in order to analyze the hierarchical structure in financial markets. Further studies have been conducted by e.g. Bonanno \cite{bonanno2004networks} or Naylor \cite{naylor2007topology} who investigated foreign exchange markets rather than stock markets. Onnela \cite{onnela2003dynamics, onnela2003dynamic} introduced the framework of a dynamically evolving MST. 

We take this concept as a starting point and move on in the following direction: Financial time series exhibit nonlinearities and we aim to capture these effects when analyzing correlation networks. Thus, we construct our networks not only based on linear Pearson correlation but based on mutual information which is sensitive to both linear and nonlinear interrelationships. Mutual information has been studied as a measure for mutual dependencies in financial time series by e.g. Dionisio \cite{dionisio2004mutual}, Fedora \cite{fiedor2014networks} and Darbellay \cite{darbellay2000entropy}. However, a detailed comparison of the properties of linear and nonlinear correlations in financial time series has not yet been done.

In this paper, we show that substantial information is lost by using a purely linear measure and propose an alternative approach by choosing mutual information as a measure that captures both linear and nonlinear correlations. The use of surrogate data \cite{theiler1992testing} allows us to create time series with the linear properties conserved but all the nonlinear properties destroyed. Thus we can compare network-topological measures based on the original and on the linearized surrogate time series and investigate nonlinear dependencies. Furthermore, this enables us to directly quantify the nonlinear correlations and derive a quantitative measure for them. While many studies have investigated financial crises from an econophysical perspective (e.g. Ref. \cite{sornette1997large, heiberger2014stock, sandoval2012correlation}) we will specifically work out the strength and influence of nonlinear correlations during crises. In order to gain useful information about the collective dynamics of the assets under study, we create networks and apply different measures such as centrality, normalized tree length and mean occupation layer \cite{onnela2003dynamics}. Finally, we apply the methods to portfolio construction: An investment strategy will be presented that takes into account nonlinear correlations in order to scale the investment exposure, which leads to a significant outperformance than compared to a fully invested portfolio.

The article is organised as follows: Section~\ref{sec:methods} introduces the data and methods used in our study. Section~\ref{sec:dependency} shows the analysis of the dependency matrices obtained from both Pearson correlation and mutual information. In section~\ref{sec:networks} we present the main results obtained from studying networks while we apply our methods to portfolio construction in section~\ref{sec:portfolio}. Our summary and the conclusions are given in section~\ref{sec:summary}.

\section{Data, Similarity Measures, Complex Networks and Surrogates}
\label{sec:methods}

\subsection{Data}
As in Onnela \cite{onnela2003dynamics} we consider the U.S. stock market. We choose a subset of stocks from the \textit{S\&P500 Index} which represents the 500 highest capitalized and thus most influential companies in the U.S.. Starting from \textit{January 2nd 1985} our data consists of the daily closing prices of all stocks that ``survived`` in the index until \textit{December 31st 2015} in order to have a consistent stock universe during the whole period. This comes to a total of $N=152$ time series with 7816 data points each. As usual, the stock prices $p$ have been converted to logarithmic returns $x$
\begin{eqnarray}
x_{i,t} = \log{p_{i,t}} - \log{p_{i,t-1}}  \ .
\label{eq:ret}
\end{eqnarray} 
In order to obtain dynamically evolving results, we divide the data in a number of overlapping windows and calculate our measure for each of the windows. Similar to Ref. \cite{onnela2003dynamics} we select a fixed-size sliding window of T = 1000 trading days which is equivalent to approximately four years of data. The step size between two consecutive windows is $\delta T= 20$ trading days. This ensures a sufficient amount of data for the calculation of the mutual information. The time series of our stocks are then defined as $X_i = \{x_{i,1} ... x_{i,T}\}$. The data used is publicly available through \textit{yahoo finance} \cite{yahoo}.

Our time horizon is long enough to investigate a number of key market events. The data covers Black Monday (October 19, 1987) when stock markets around the world crashed for the first time after World War II. From 1997 to 2001 the markets were subject to excessive speculation and overvaluation of many technology companies which led to the Dot-com bubble. The bubble bursted during 2002 with significant declines taking place in July and September. Finally, our data includes the 2007/2008 subprime mortgage crises. At that time the market declined after its all time high in October 2007 and a crash occurred after the collapse of Lehman Brothers on September 15, 2008. In addition, our considered time span also includes a number of major global political events. These include the fall of the Berlin wall on November 9, 1989, which triggered the collapse of the Soviet Union as well as the 9/11 attacks on September 11, 2001.

\subsection{\label{sec:level2}Measures for Mutual Dependencies}
\label{sec:measures}

\subsubsection{Pearson Correlation Coefficient}
The standard approach in finance to quantify mutual dependencies between stocks is the Pearson correlation coefficient $\rho$,
\begin{eqnarray}
\rho_{X_i,X_j}=\frac{\sum^{n}_{t=1}(x_{i,t}- \bar{x}_i)(x_{j,t}-\bar{x}_j)}{\sqrt{\sum^{n}_{t=1}(x_{i,t}- \bar{x}_i)^2} \sqrt{\sum^{n}_{t=1}(x_{j,t}-\bar{x}_j)^2}} \ ,
\label{eq:one}
\end{eqnarray} \\ 
where $x_i$ are the stock returns of stock $i$ and $\bar{x}_i$ their mean, respectively. It is bound to the interval [-1,1] and allows to directly compare correlations of different asset pairs as it is a normalized measure. A serious problem with Pearson correlation, however, is that it only captures linear interrelationships. 
\begin{figure}[t] 
  \begin{center}
     \includegraphics[width=1\linewidth]{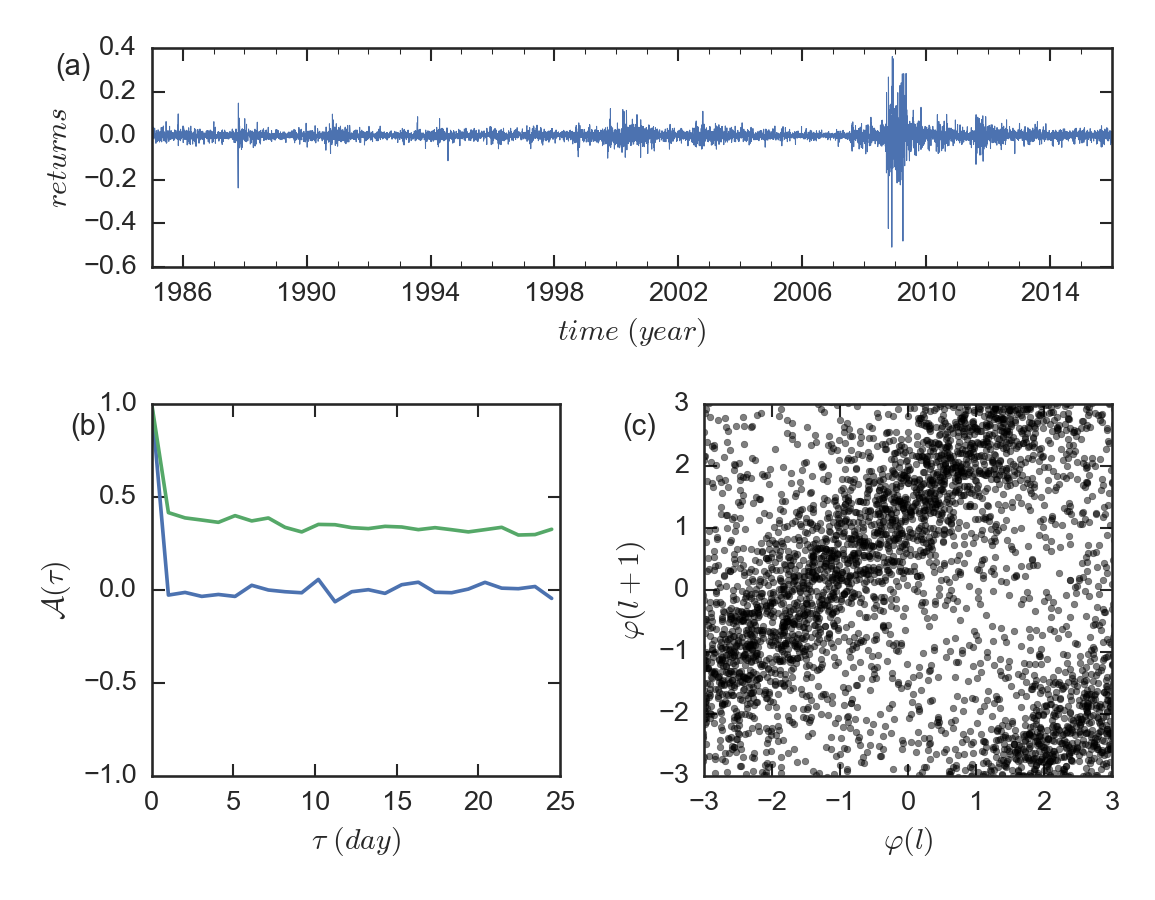}
  \caption{(a) Return time series, (b) autocorrelation function of returns (blue) and absolute values of returns (green), (c) phase map of Lincoln National Corporation (LNC) stock returns. Phase map: Phases $\varphi(l)$ are scattered against neighboring phases $\varphi(l+1)$.}
  \label{fig:phases}
  \end{center}
\end{figure}
\subsubsection{Mutual Information}
It is well known that financial time series exhibit nonlinear effects \cite{hsieh1995nonlinear}. This is exemplified in Fig. \ref{fig:phases}. There we show the time series, autocorrelation function and phase map of the Lincoln National Corporation (LNC) stock. Phase maps are sets of points $G = \{\varphi(l), \varphi(l+\Delta)\}$ where $\varphi(l)$ is the $l^{th}$ mode of the Fourier transform
\begin{eqnarray}
\varphi(l)=\arg \sum_{t=0}^{T-1} x_t e^{-2 \pi i t l/T}
\label{eq:mode}
\end{eqnarray} 
and $\Delta$ a mode delay with $\Delta = 1$ in this example. A random uncorrelated distribution would lead to a random distribution of points in the phase map. Here, we observe a significant stripe pattern. This clearly indicates the presence of nonlinear effects in our stock returns data \cite{rath2015time}. 
R\"ath \emph{et al.} \cite{rath2012revisiting} already showed the presence of such stripe pattern in the data of the Dow Jones index. As we observed similar effects not only for the LNC example, but also for the other stocks in our dataset we conclude that phase correlations are a generic feature. Furthermore, the autocorrelation of the returns immediately drops to zero which means that the time series does not exhibit a linear memory. At the same time the autocorrelation of the absolute values of the returns does not drop to zero. There is no linear process that can generate a behavior like this \cite{hsieh1995nonlinear}.

The consequence is the following: If there are nonlinear effects present in financial time series, the purely linear Pearson correlation coefficient captures only a fraction of mutual dependencies and thus a significant amount of information is ignored. Hence it would be beneficial to use a different measure that captures all kind of relationships between two time series. 

An appropriate solution to this problem is the use of \textit{mutual information} $\widetilde{I}(X_i,Y_j)$ \cite{kraskov2004estimating} as a measure for mutual dependencies
{\small
\begin{eqnarray}
\widetilde{I}(X_i,X_j)= \int_{}\int_{}\ p(x_i,x_j)\ log\bigg(\frac{p(x_i,x_j)}{p(x_i) p(x_j)}\bigg)dx_i dx_j \ ,
\label{eq:two}
\end{eqnarray} 
}where $p(x_i,x_j)$ is the joint probability density function and $p(x_i),\ p(x_j)$ the marginal \textit{PDFs} respectively. This is because mutual information is sensitive to both linear and nonlinear correlations. Alternatively one can express mutual information through the marginal- and joint entropies of the two variables
\begin{eqnarray}
\widetilde{I}(X_i,X_j)=H(X_i) + H(X_j) - H(X_i,X_j) \ .
\label{eq:three}
\end{eqnarray} 
$H(X_i)$ denotes the entropy of variable $X_i$ and is defined as
\begin{eqnarray}
H(X_i) =  - \sum^{}_{x_i} \ p(x_i) \ log(p(x_i)) \ ,
\label{eq:four}
\end{eqnarray} 
while the joint entropy $H(X_i,X_j)$ of variables $X_i$ and $X_j$ reads
\begin{eqnarray}
H(X_i,X_j) =  - \sum^{}_{x_i} \sum^{}_{x_j} \ p(x_i, x_j) \ log(p(x_i, x_j)) \ .
\label{eq:five}
\end{eqnarray} 
We switched here to the discrete formulation using sums instead of integrals. This has a very important reason: Mutual Information is not a normalised measure and thus it can take on values between zero and \textit{infinity}. For discrete variables one can normalize mutual information \cite{strehl2002cluster} in the interval [0,1] where 0 means that both variables do not share any information and 1 means completely identical probability distributions
\begin{eqnarray}
I(X_i,X_j)=\frac{\widetilde{I}(X_i,X_j)}{\sqrt{H(X_i)H(X_j)}} \ .
\label{eq:six}
\end{eqnarray} 
We use binning methods to estimate the probability density function in order to ensure normalizability. Heuristically we find $\lceil{\sqrt{T/4}}\ \rceil{}$ to be a good choice for the number of bins which leads to 16 in our case of window size $T=1000$. We are aware that alternative methods like kernel density or nearest neighbour based estimators \cite{kraskov2004estimating} would give a better approximation of the probability distributions. However, this would lead to the problem of not normalizable mutual information. 

\subsection{\label{sec:level2}Network Construction}
As first proposed in Mantegna \cite{mantegna1999hierarchical} the concept of graphs and more specifically Minimum Spanning Trees is very useful in order to summarize the vast amount of information stored in correlation matrices. We choose the MST as the main type of network to analyze due to its simplicity. Here the concept is to connect \textit{N} nodes (assets) by \textit{N-1} edges under the constraint that distances are minimal and hence dependencies maximal. We construct MSTs by using Prim's algorithm \cite{prim1957shortest}. The advantage of the MST in comparison to other types of networks is that we do not need to choose any parameters. Instead, it emerges automatically and thus ensures comparability. Using the two different measures from section~\ref{sec:measures}, we aim to investigate the dynamical evolution \cite{onnela2003dynamics} of graphs that capture only the linear or both the linear and nonlinear correlations. The results are then compared and we analyze during which market periods we observe differences between the measures.
\begin{figure*}[t!] 
  \begin{center}
     \includegraphics[width=1\linewidth]{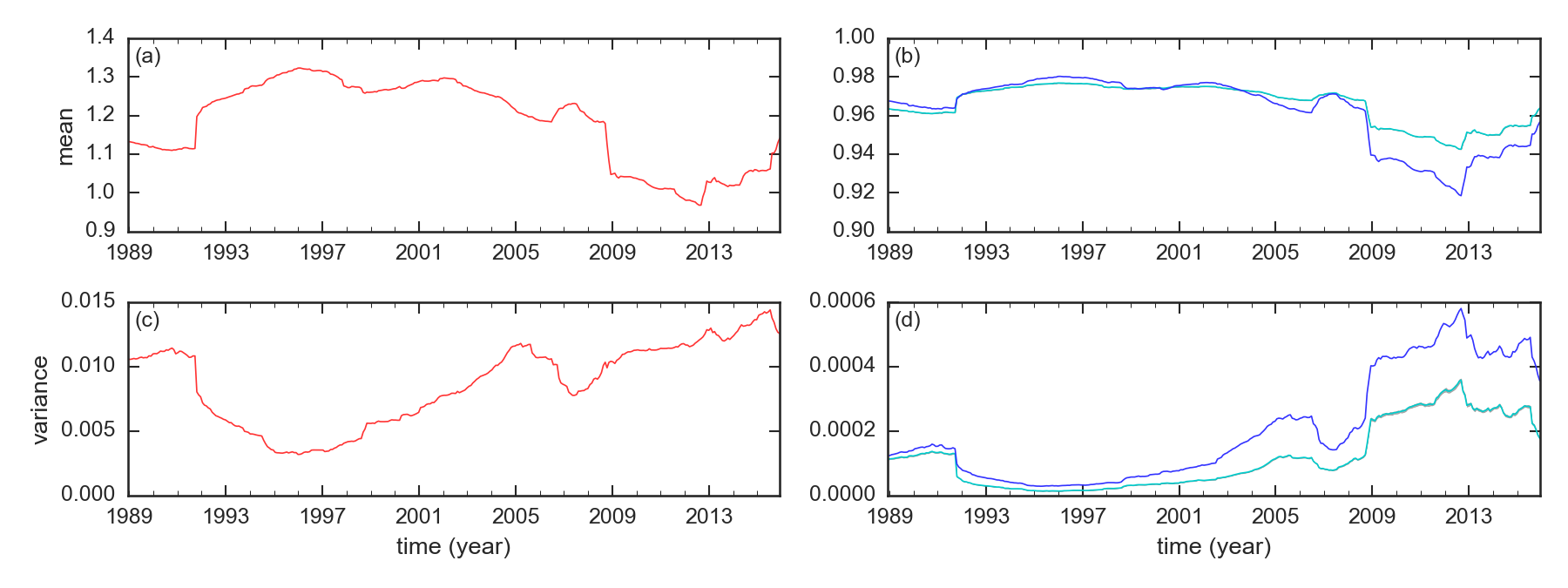}
  \caption{Left: Mean (a) and variance (c) of the coefficients of the Pearson correlation based distance matrix calculated in each time step. Right: Mean (b) and variance (d) based on the mutual information of the original series (blue) and the average distance matrix of all surrogate realizations (cyan). Plus/minus one sigma error ranges (grey) are drawn for the surrogate realizations, however, errors are so small that one can barely seen them.}
  \label{fig:meanvar}
  \end{center}
\end{figure*}

Before we are able to construct networks we need to convert the correlation and mutual information matrices to distance matrices in each time step $t$ by using an appropriate metric. For mutual information this is straightforward by applying 
\begin{eqnarray}
d^{MI,t}_{X_i,X_j} = 1 - I^t(X_i,X_j) 
\label{eq:midistance}
\end{eqnarray} 
to the normalized mutual information. 

As Pearson's correlation coefficient can also take on negative values, we have to transform it to a non-negative distance measure. For this we use the transformation 
\begin{eqnarray}
d^{corr,t}_{X_i,X_j} = \sqrt{2(1-\rho^t_{X_i,X_j})} 
\label{eq:distance}
\end{eqnarray} 
as discussed in Onnela \cite{onnela2003dynamics} which fulfills all requirements for a distance metric. When comparing Eq.~(\ref{eq:midistance}) and Eq.~(\ref{eq:distance}) we observe that both distance metrics behave differently if time series are linearly anti-correlated, i.e. when their Pearson correlation coefficient is negative. However, in our case all stock time series are positively correlated in terms of Pearson correlation, which is typically the case in stock markets. If there were negatively correlated time series, one could avoid this problem by e.g. taking the absolute value of the Pearson correlation when calculating the distance. We can now construct MSTs $\pmb{T}^t$ in each time step $t$ based on the two dependency measures. 

In addition, we also construct so called threshold networks where we only keep connections between nodes with a distance less than a certain threshold. Heuristically we find a threshold of 20\% to be an appropriate value i.e. connecting assets with the smallest 20\% of distances and thus highest correlations. 

\subsection{\label{sec:level2}Surrogates}
When using \textit{mutual information} as a measure for mutual dependencies we capture both linear and nonlinear correlations. In order to analyze the effects that are due to nonlinear dependencies we need to separate linear and nonlinear contributions.

Surrogate data allow us to exactly achieve this separation by destroying nonlinear effects of the time series while keeping all linear properties \cite{theiler1992testing}. In this study we use so called \textit{Fourier transform (FT) surrogates} where we Fourier transform the time series and thus separate all linear properties into the amplitudes while the nonlinear properties are stored in the phases. By adding uniformly distributed random numbers to the Fourier phases we destroy all nonlinear properties while the linear ones stay untouched. An inverse Fourier transformation gives us then the final surrogate data
\begin{eqnarray}
x^*_k(t)= \mathcal{F}^{-1}\{\widetilde{X}(f)\} = \mathcal{F}^{-1}\{X(f)e^{i\phi_k(f)}\} \ .
\label{eq:surro}
\end{eqnarray} 
Equation ~(\ref{eq:surro}) defines the \emph{k}-th surrogate realization. We create $K=20$ realizations and average over the measures calculated for each realization in order to get a more stable result. $\mathcal{F}^{-1}$ denotes the inverse Fourier transform operator and $e^{i\phi_k(f)}$ the \textit{k-th} set of uniformly distributed random phases that is added to the Fourier transform $X(f)$ of the original time series \textit{X}. Prichard and Theiler \cite{prichard1994generating} showed that it is also possible to conserve the Pearson correlations by adding the same set of random numbers onto the phases of all time series. This is because the Fourier transformed Pearson correlation depends only on the phase differences between time series, which is then unaffected. 

We would like to point out that we do not do a rank-ordered remapping of the data onto a Gaussian distribution at the beginning of the procedure. Thus we test for static and dynamic nonlinearities at the same time. Converting prices to logarithmic returns could potentially induce static nonlinearities. We convinced ourselves that the results presented later would not change much after performing the above mentioned remapping. Thus, we conclude that the results are mainly driven by dynamic nonlinearities.

\section{Analysis of dependency matrices}
\label{sec:dependency}
\subsection{\label{sec:level2}Distance Matrix Coefficients}

Before constructing networks we first examine the dynamical evolution of the distribution of the distance matrix coefficients based on Pearson correlation and mutual information. In case of the surrogates we average over all $k$ realizations 
\begin{eqnarray}
\mathcal{M}^m_{Surro} = \langle  \mathcal{M}^m\{d^{MI^{*}}_k\} \rangle_k \ ,  
\label{eq:surroavg}
\end{eqnarray} 
where $\mathcal{M}^m$ is the \emph{m}-th moment of the distribution of the coefficients from the distance matrix $d^{MI^{*}}_k$ obtained from the \emph{k}-th surrogate realization. 
\begin{figure}[b!] 
  \centering
     \includegraphics[width=0.52\textwidth]{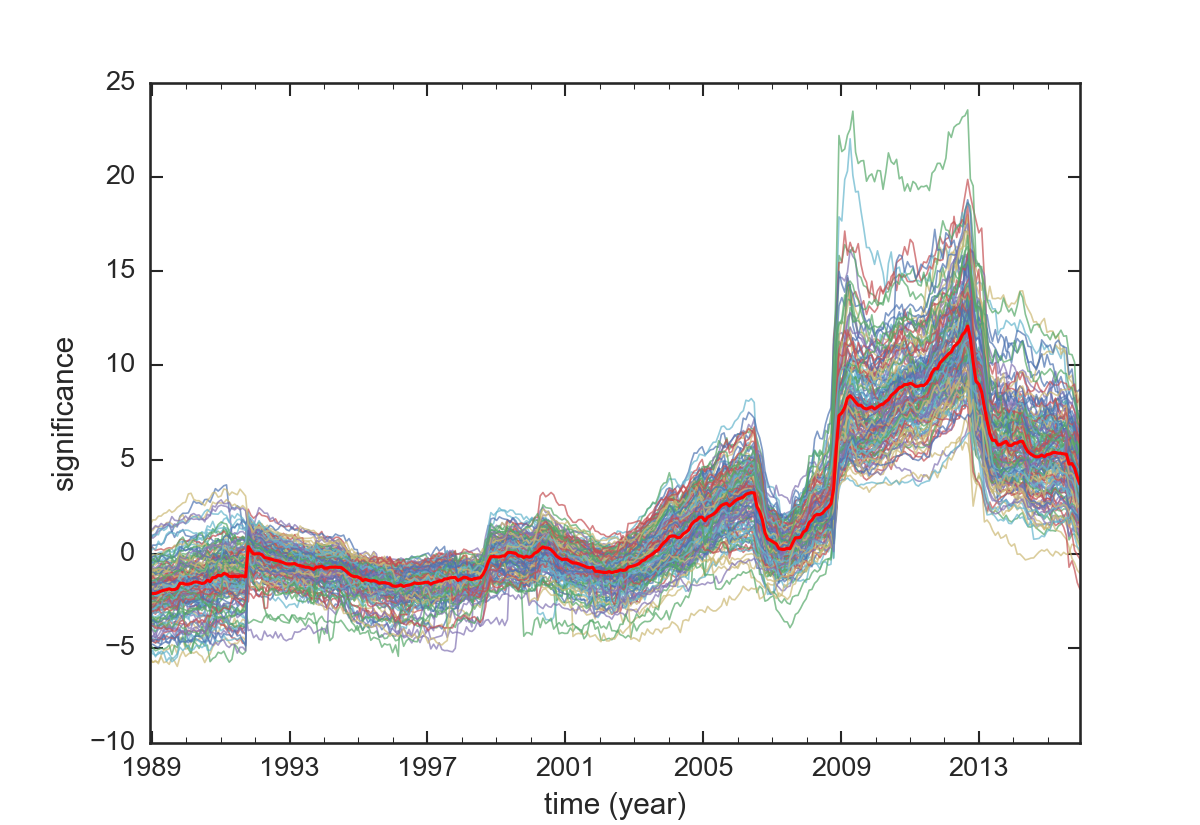}
  \caption{Average significance coefficient $\overline{\chi_{sig}(X_i)}$ of each asset - 
  Red line: Global average over all individual assets.}
  \label{fig:sig}
\end{figure}

Figure~(\ref{fig:meanvar}) shows the dynamically evolving mean and variance of the distance matrices based on Pearson correlation and mutual information. As we expect, the resulting moments based on the Pearson correlation of the original and surrogate time series are exactly identical per construction. For mutual information, however, we note that the results of the surrogates are more similar to the Pearson correlation based results. The mean mutual information based distance of both the original time series and the surrogate series evolves very similarly until the 2008 financial crisis. However, starting from November 2008 the mean based on the original time series becomes lower than the surrogate based average. Since less distance means higher average mutual information we conclude that nonlinear effects lead to stronger dependencies among the time series triggered by the 2008 financial crisis. The strength of nonlinear correlations further grows throughout the European debt crisis until reaching its peak in the middle of 2012. This result is a little surprising as it disagrees with the expectation that interdependencies reduce mainly to linear correlations during crises.
It is interesting to notice that the variance of the distance matrix coefficients (lower row of Fig.~(\ref{fig:meanvar})) behaves slightly different. Nonlinear effects are increasing the variance starting from the onset of the Dot-com bubble in 1998 and amplify throughout the whole remaining period. Taking into account both moments of the distance matrices we can already see that nonlinear effects are clearly present even at a very general level of analysis. 
\begin{figure}[b!] 
  \centering
     \includegraphics[width=0.50\textwidth]{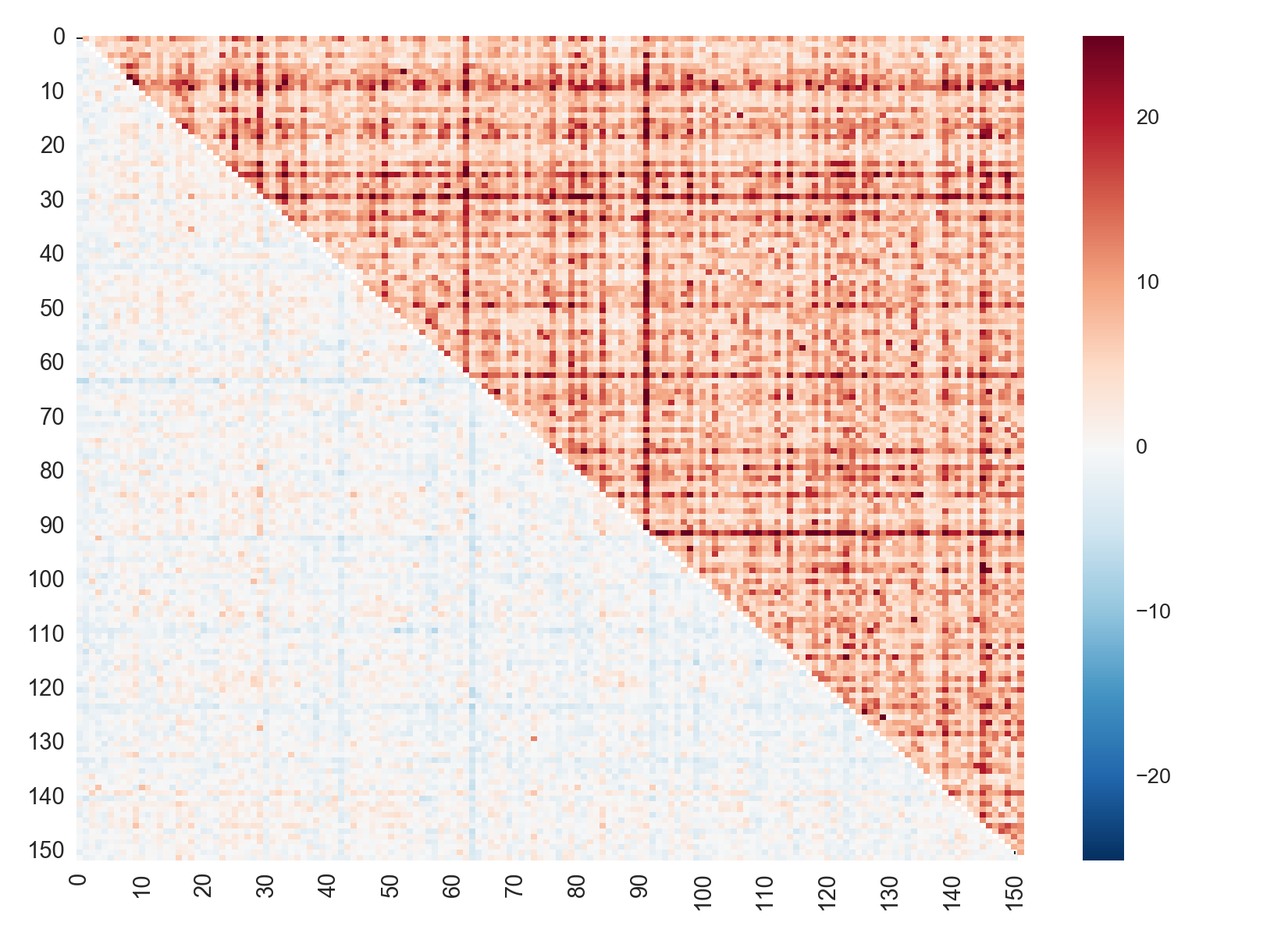}
  \caption{Significance matrices $\pmb{\chi}_{sig}$; Lower half: Calm market environment in November 1998, Upper half: Turbulent market environment in May 2009 after the crash triggered by the Lehman collapse on September 15, 2008. Labels denote stock indices.}
  \label{fig:sigMatrix}
\end{figure}

\subsection{\label{sec:level2}Deriving Nonlinear Dependencies}

In order to explicitly express the significance of nonlinear correlations we derive a significance measure $\chi_{sig}$ by first calculating the mutual information matrix of all surrogate realizations. We then take the average over all surrogate realizations and subtract it from the mutual information matrix of the original time series
\begin{eqnarray}
\chi_{sig}(X_i,X_j)= \frac{I(X_i,X_j) - \langle I(X^*_{i,k},X^*_{j,k}) \rangle_k}{\sigma_{I^*}} \ .
\label{eq:nine}
\end{eqnarray} 
Finally, we normalize it by the standard deviation of the surrogate realizations $\sigma_{I^*}$. Figure~\ref{fig:sig} shows the time evolution of the column averages of the significance matrix, which represent the significance of the mean nonlinear interaction of one stock with all others
\begin{eqnarray}
\overline{\chi_{sig}(X_i)}= \langle \chi_{sig}(X_i,X_j) \rangle_j \ .
\label{eq:sigavg}
\end{eqnarray} 
It clearly shows that the 2008 crisis triggers strong nonlinear effects, which we do not see during the early 90s recession or other turbulent market phases like the Dot-com crisis. The period of uncertainty starting in the end of 2009 caused by the European sovereign debt crisis further amplifies the significance of nonlinear interactions and only slowly declines towards the present. The highest significance values are achieved by Lincoln National Corporation (insurance and investments) and Citigroup (financial services) in May 2009 shortly after stock prices bottomed in the course of the financial crisis.
  
By taking the absolute value of the difference between mutual information of the original data and the average of the surrogate realizations and dividing it by the original mutual information we define a measure for the strength of overall nonlinear correlations
\begin{eqnarray}
\zeta_{nlc}(X_i,X_j)= \frac{\left | I(X_i,X_j) - \langle I(X^*_{i,k},X^*_{j,k}) \rangle_k \right |}{I(X_i,X_j)} \ ,
\label{eq:ten}
\end{eqnarray} 
which we will later use in the portfolio optimization section. It tells us, which amount of the overall mutual information is due to nonlinear mutual dependencies. However, it is not entirely clear what the meaning of ``negative nonlinearities" would be, i.e. when the average mutual information of all surrogate realizations takes a higher value than the original one. This would correspond to saying that after destroying nonlinear effects, both time series share more information than before. We see in Fig.~\ref{fig:sig} that there are certain periods in time where the mutual dependencies of original and surrogate time series are almost identical and hence only linear correlations play a role. The lower half of Fig.~\ref{fig:sigMatrix} shows the significance matrix at a rather calm market period in November 1998. In contrast, the upper half shows significantly higher significance values in the aftermath of the 2008 financial crisis in May 2009. Darker red stripes indicate particularly strong nonlinear correlations of one asset with all others as for e.g. Lincoln National Corporation (LNC), Citigroup (C) and General Electric (GE).
\section{Network-Based Analysis}
\begin{figure}[b!] 
  \centering
     \includegraphics[width=1\linewidth]{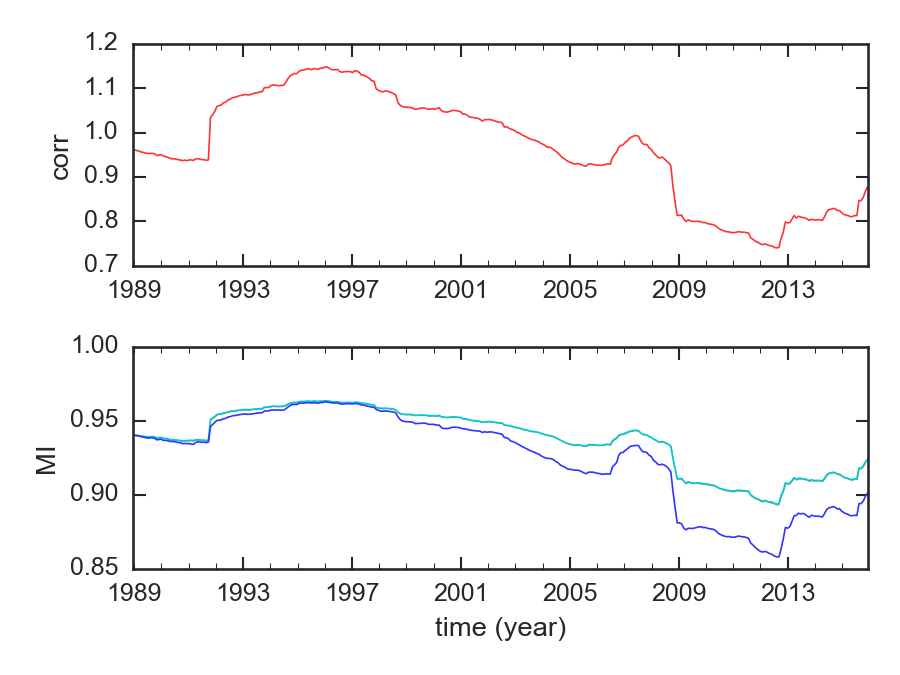}
  \caption{Top: Normalized Tree Length of Pearson correlation based MST. Bottom: Same for mutual information of original data (blue) and surrogates (cyan).}
  \label{fig:normtree}
\end{figure}
\begin{figure*}[t!] 
  \centering
     \includegraphics[width=1\linewidth]{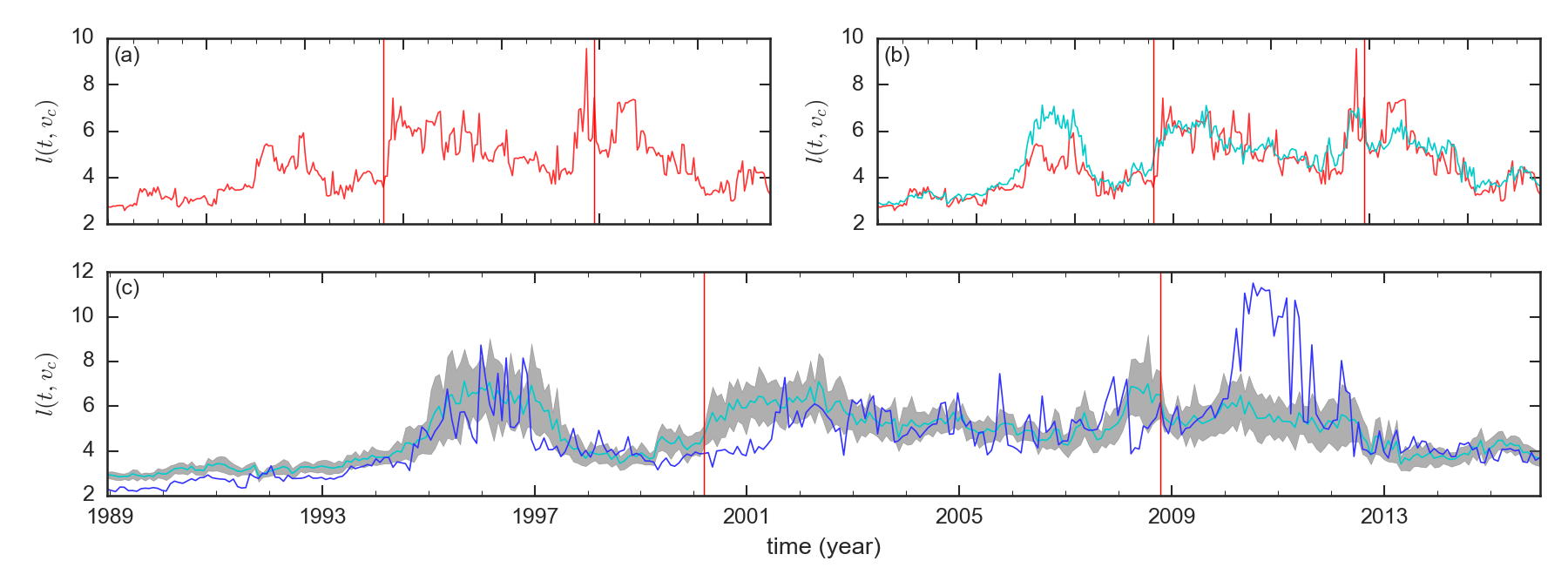}
  \caption{(a) Dynamic mean occupation layer of Pearson correlation based MST. (b) Mutual Information of surrogate series (cyan) vs. Pearson correlation based result (red). (c) Same for mutual information of original series (blue) and surrogates (cyan) with plus/minus one sigma error range in grey. Vertical lines indicate the bursting of the Dot-com bubble in March 2000 and the acceleration of the crash after the Lehman Brothers collapse on September 15, 2008.}
  \label{fig:dynlayer}
\end{figure*}
\label{sec:networks}
\subsubsection{\label{sec:level2}Normalized Tree Length}
As the next step, MSTs have been constructed from the Pearson correlation and mutual information based distance matrices using Prim's algorithm \cite{prim1957shortest}. First of all, we take a look at the normalized tree length \cite{onnela2003dynamics} in Fig.~\ref{fig:normtree} which is defined as 
\begin{eqnarray}
L(t)= \frac{1}{N-1} \sum^{}_{d^t_{X_i,X_j} \in \ \pmb{T}^t} d^t_{X_i,X_j} \ ,
\label{eq:sigavg}
\end{eqnarray} 
where $t$ denotes the time step in which the MST has been constructed. We notice that the qualitative behavior is very similar to the mean of the distance matrix elements. However, in the case of mutual information, there is a gap emerging between original time series and surrogate based trees already in the middle of 1998. This is when the Dot-com bubble slowly started growing. This happens earlier than compared to the average of all distance elements in Fig.~\ref{fig:meanvar}. 
\subsubsection{\label{sec:level2}Mean Occupation Layer}

Another interesting property of the networks we attend to from Ref. \cite{onnela2003dynamics} is the dynamic mean occupation layer
\begin{eqnarray}
l(t,v_c)= \frac{1}{N} \sum^{N}_{i=1}  \mathcal{L}(v^t_i) \ ,
\label{eq:occup}
\end{eqnarray}
where $\mathcal{L}(v^t_i)$ denotes the level of node $v_i$ at time step $t$. The level $\mathcal{L}(v^t_i)$ measures the distance of node $v_i$ from the central vertex in terms of absolute numbers of edges. Thus the mean occupation layer reflects the average distance of nodes from the center of the network. The center is dynamically determined in each time step based on degree centrality which we are explaining in the following section. We can interpret the mean occupation layer as a measure for the diversification potential within the set of our stocks. 
As shown in Fig.~\ref{fig:dynlayer}, the result based on the linear measure (a) indicates that the mean distance from the center of the network increases during financial bubbles. It peaks in June 2008 and thus shortly before the crash occurred as well as during the Dot-com bubble in 2001. 
Figure~\ref{fig:dynlayer} (b) demonstrates that the results based on Pearson correlation and mutual information of the linearized surrogate time series evolve very similarly. However, the lower plot (c) where we compare the mutual information based result between the original and surrogate time series exhibits the following interesting features. During the Dot-com bubble and its subsequent crash the surrogate data yields a higher mean occupation layer as compared to the results based on the original time series. This means that linear correlations lead to stronger diversification and thus we could state that linear correlations are dominating the topology of the network during the Dot-com crash. In contrast, the behavior during the 2008 financial crisis is different. The original time series mean occupation layer drops below the surrogate layer in the beginning of 2008. However, the sharp decline in stock prices after the Lehman collapse in September 2007 triggered a substantial increase in the original time series layer. It further grows during the aftermath of the crisis and peaks in the middle of 2010. Here it reaches a level of around 11.5 which is almost twice as high as the level of the surrogate based layer and thus indicates that the nonlinear tree diversifies significantly stronger. 
This shows that the 2008 financial crisis and the Dot-com crash have a different character, as we did not observe similar effects during the latter. Further support for this observation is provided by Fig.~\ref{fig:sig} where we see that nonlinear correlations are weak during the Dot-com period but strongly grow starting in early 2008 - before the crash occurred. We observe similar effects later on in section $V$ where we apply a portfolio optimization strategy, which is based on nonlinear correlations.
\subsubsection{\label{sec:level2}Centrality}
\begin{figure}[t!] 
 \begin{center}
     \includegraphics[width=1\linewidth]{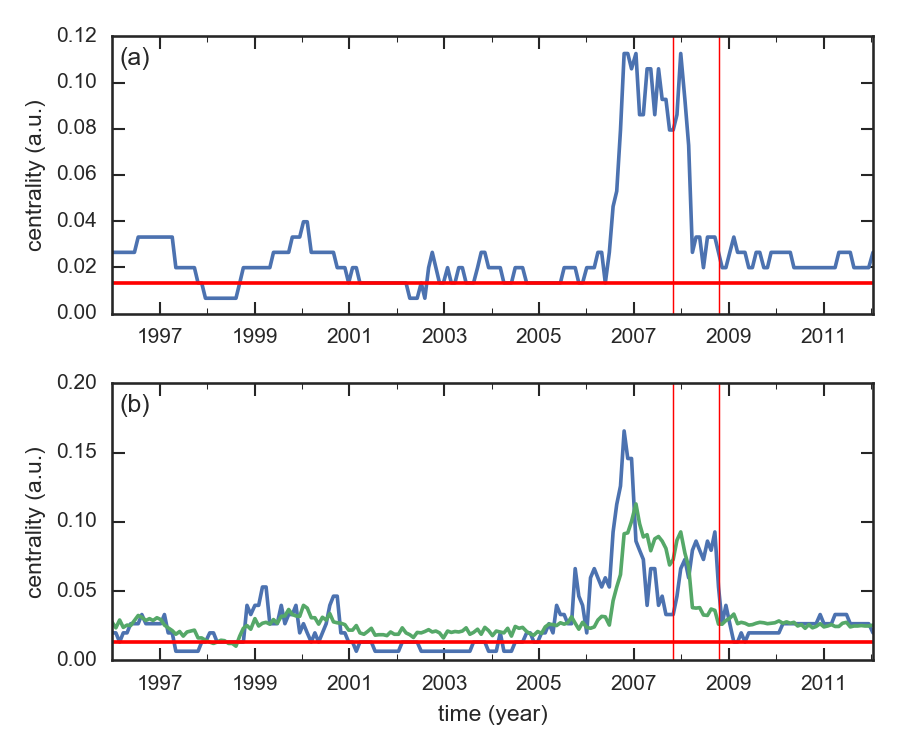}
  \caption{Degree centrality of JPM based on Pearson correlation (a) and mutual information (b) (blue: original data, green: surrogate data) vs. network avg. (red). The vertical lines indicate the peak of the stock prices on October 15, 2007 after which they began falling as well as the acceleration of the crash after Lehman Brothers collapsed on September 15, 2008.}
  \label{fig:jpm}
  \end{center}
\end{figure}

\begin{figure}[t!] 
 \begin{center}
     \includegraphics[width=1\linewidth]{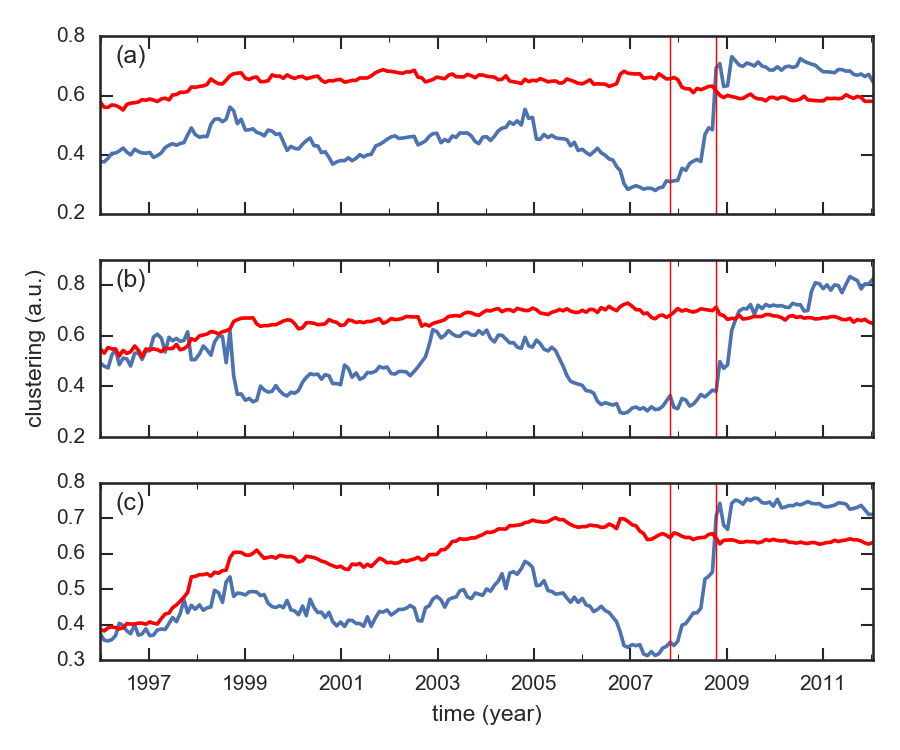}
  \caption{Clustering coefficient of JPM (blue) vs. network avg. (red) for Pearson correlation (a), original mutual information (b) and surrogate mutual information (c).}
  \label{fig:clustering}
  \end{center}
\end{figure}

We use the concept of degree centrality to determine how central and thus important a stock is in our networks. Degree centrality 
\begin{eqnarray}
deg_{i}(t)= \frac{1}{N} \sum^{N}_{j=1}  \mathcal{D}(d^t_{X_i,X_j})
\label{eq:degree}
\end{eqnarray} 
\begin{figure}[b!] 
  \centering
     \includegraphics[width=0.50\textwidth]{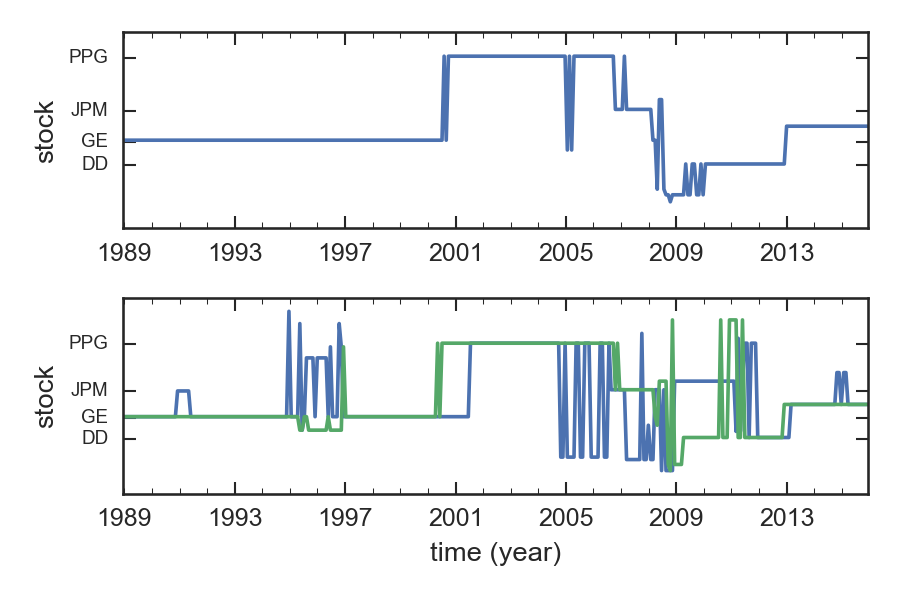}
  \caption{Assets with the highest degree centrality. Top: Pearson correlation based; Bottom: Original mutual information (blue) and surrogate mutual information (green) based.}
  \label{fig:degree}
\end{figure}of stock $i$ simply counts the number of connections to other stocks where $\mathcal{D}(d^t_{X_i,X_j}) = 1 $ if $d^t_{X_i,X_j} > 0 $ which mean that stocks $X_i$ and $X_j$ are connected and 0 else. In Fig.~\ref{fig:degree} we can see a dynamical overview of the most central stocks in our networks. In general, the dominating assets in terms of degree centrality are very similar in both correlation and mutual information based networks. However, the results based on mutual information show more fluctuations especially from 1994 to 1997 and from 2005 until 2012.
Until 2000, General Electric (GE) was by far the most central element. This could be driven by GE's important role in the US economy being a broadly diversified company and one of the largest employers. Furthermore, GE's large market capitalization was further increasing until the second half of 2000 (Source: Bloomberg). However, in the course of the Dot-com bubble and its subsequent crash from 2000 until 2002 GE's market capitalization rapidly dropped and at the same time it's importance in the network in terms of degree centrality. Instead, PPG Industries -- a chemicals and specialty materials supplier -- emerged as the most central node in both correlation and mutual information based networks lasting until late 2006. From then on J.P.Morgan (JPM) emerged as the most central node in all networks, which is particularly interesting in the course of the 2008 financial crisis. 
JPM's centrality grew excessively and peaked in late 2006 as shown in Fig.~\ref{fig:jpm} --- long before the subprime bubble started bursting in October 2007. In the course of the stock market crash the centrality then rapidly declined towards the global network average in early 2009. In the following analysis we will focus on the time between 1998 and 2010 since the most interesting phenomena occur here. In order to gain additional insights about the role of JPM we also construct a threshold network where we simply connect the pairs of nodes with the highest 20\% of dependency coefficients. There we observe similar effects in the clustering coefficient of JPM during the same periods. The clustering coefficient
\begin{eqnarray}
c_{i}(t)= \frac{1}{deg_{i}(t) (deg_{i}(t) -1)} \sum^{}_{jk} (\tilde{w}_{ij} \tilde{w}_{jk} \tilde{w}_{ik})^{1/3} 
\label{eq:clus}
\end{eqnarray} 
describes the transitivity of JPM where $\tilde{w}_{ij} = w_{ij} / max(w) $ denotes normalized edge weight of node $i$ and $j$ \cite{saramaki2007generalizations}. A high clustering coefficient means that neighbours of JPM are highly connected while a low clustering coefficient means that they tend to not have connections. As shown in Fig.~\ref{fig:clustering}, the clustering coefficient of JPM starts moving away from the global network average in 2005 and declines until mid of 2007. Together with the centrality increasing during the same time period we can interpret this the following way: During the growth of the subprime bubble, JPM as America's largest financial institution takes on an increasingly important role in the network. At the same time the clustering coefficient decreases and indicates that there are fewer connection between JPM's neighbours making JPM the ``driving force" in the network. More interestingly, all this happens long before the crash occurs with strong deviations from the average network state emerging very early. This could mean that JPM acts as kind of an early warning system, which is signaling through centrality and clustering measures that the financial markets in the US show abnormal behavior. Furthermore, we found that the same effects occurred during the Dot-com bubble and its subsequent crash around 2001/2002 and somewhat weaker during the early 90s recession (results not shown).  \\ 
\begin{table*}
\begin{ruledtabular}
\begin{tabular}{lllll}
 $Measure\ Base$&$Measure\ \mathcal{M}$& $I(original)$&$\langle I(surrogate) \rangle$ &$\sigma(I(surrogate))$     \\ \hline  \noalign{\vskip 0.05in}    
$Distance\ Matrix$&Mean&$0.958$ &$0.992$ &$4.50 \times 10^{-5}$ \\
$Distance\ Matrix$&Variance&$0.839$&$0.845$ & $5.29 \times 10^{-4}$\\
$Distance\ Matrix$&Skewness&$0.925$&$0.947$  & $1.04 \times 10^{-3}$\\
$Distance\ Matrix$&Kurtosis&$0.886$&$0.891$  & $2.34 \times 10^{-3}$ \\
$Minimum\ Spanning\ Tree$&Normalized Tree Length&$0.977 $&$0.992$   & $7.78 \times 10^{-5}$ \\
$Minimum\ Spanning\ Tree$&Betweenness Centrality&$0.767$&$0.908$ & $0.017$\\
$Minimum\ Spanning\ Tree$&Mean Occupation Layer&$0.594$&$0.830$ & $0.033$\\
\end{tabular}
\caption{In each time step we calculate the above measures based on Pearson correlation as well as mutual information of the original and surrogate time series. Here we show the Pearson correlation coefficients of the resulting set of values based on linear Pearson correlation with: 1. The results based on the mutual information of the original data denoted by $I(original)$ 2. The results based on the mutual information of the surrogate data denoted by $\langle I(surrogate) \rangle$ where we averaged over the results of all surrogate realizations. $\sigma(I(surrogate))$ is their standard deviation respectively.}
\label{table:overview}
\end{ruledtabular}
\end{table*}
When comparing the Pearson correlation based measures to the mutual information based measures we observe that in the latter case the effects are qualitatively stronger articulated. In Fig.~\ref{fig:jpm} (b) we present the mutual information based results for the degree centrality. Compared to the measures based on the original time series, the quantitative strength of the degree centrality is significantly lower for the surrogate data. Moreover, both centrality measures begin their most significant drop in October 2007, which is exactly when the stock markets started moving downwards after peaking. However, centrality measures of the networks constructed from the mutual information based on the original time series are rapidly falling after their peak in October 2006 and thus around one year before the stock market crisis started. When comparing the results of the original time series to the surrogate results we observe that there is a sharp decline from the centrality peak in October 2006 until October 2007, which is only happening in the measures for the original series. This indicates that JPM's centrality is influenced by strong nonlinear effect during this period. From October 2007 until September 2008 centrality grows again --- this corresponds to the period where stocks started to fall after their all time high in October 2007. When stock prices drop even faster in September 2008 after the investment bank Lehman Brothers went bankrupt, JPMorgan's centrality suddenly decreases as well. We would like to mention that we observed similar behavior for Goldman Sachs and Lehman Brothers in a different set of data (results not shown).            

Table~(\ref{table:overview}) summarizes different network and distance matrix measures and shows how similar both the Pearson correlation and the mutual information based results are in terms of the Pearson correlation coefficient of their resulting sets of values. Values close to unity signify a close similarity of the Pearson correlation and mutual information based results. We observe that the surrogate mutual information and Pearson correlation based results have a higher similarity than the original data mutual information and Pearson correlation based results. Thus there are nonlinear correlations present and surrogate data is a good method for comparison when using mutual information as a measure for mutual dependencies. 

Unlike during the periods of financial crises mentioned above, we do not observe effects during major political events such as the fall of the Berlin wall in November 1989 or the 9/11 attacks in September 2001. Neither the mean occupation layer nor the normalized tree length indicates significant changes in market correlation structure.

\section{Portfolio optimization}
\label{sec:portfolio}
\subsection{\label{sec:level2}Markowitz Portfolio Construction}
To make use of our concepts we apply them to the problem of portfolio construction. A standard approach in finance is the mean-variance optimization developed by Harry Markowitz in 1952 \cite{markowitz1952portfolio} where the variance of a portfolio is minimized given a certain target return or risk aversion factor. The expected return $\mu_P$ of the portfolio $P$ then is
\begin{eqnarray}
\mu_P = E[P(\textbf{w})]= \sum^{}_{i} w_i E[X_i],
\label{eq:expret}
\end{eqnarray} 
where $w_i$ is the weight and $E[X_i]$ denotes the expected return of asset $i$. The expected return of each asset is assumed to be its median during the historical time window $T = 1000$ after excluding outliers by applying an interquartile range based filter. This is not a very good approximation because the autocorrelation function of the returns quickly falls towards zero and hence past returns do not tell much about the future. For the sake of simplicity, however, we decided to use this standard approach. The variance of our portfolio is then given by
\begin{eqnarray}
\sigma^{2}_{P}(w)= \sum^{}_{i}  \sum^{}_{j} w_i w_j \sigma_{X_i} \sigma_{X_j} \rho_{X_i,X_j},
\label{eq:portvar}
\end{eqnarray} 
where $\sigma_{X_i}$ is the standard deviation of the returns of asset $i$ and $\rho_{X_i,X_j}$ the Pearson correlation coefficient of assets $i$ and $j$. The expression $\sigma_{X_i} \sigma_{X_j} \rho_{X_i,X_j}$ is also called covariance $\sigma_{X_i,X_j}$ while we denote the covariance matrix of all assets as $\Sigma$. The optimal portfolio described by the weight vector $\textbf{w}$ given some target return $\mu_P$ is then obtained by minimizing
\begin{eqnarray}
\textbf{w}^T \Sigma \ \textbf{w}
\label{eq:weight}
\end{eqnarray} 
subject to 
\begin{eqnarray}
\mu_P = \textbf{R}^T \textbf{w} 
\end{eqnarray} 
and 
\begin{eqnarray}
\sum^{}_{i} w_i = 1,
\end{eqnarray}
where $\textbf{R}^T$ is the vector of the expected returns of the assets. Now we are left with one more decision: Which target return should we select for our portfolio optimization? To ensure a consistent strategy we construct Markowitz optimized portfolios for all possible target returns and then choose the one where the fraction of portfolio return and volatility
$ \frac{\mu_P}{\sigma^{2}_{P}(w)}$ is the largest --- the so called \textit{maximum sharpe ratio}  portfolio. We do not allow short selling in this example which means that asset weights $w_i$ have to be positive.

\subsection{\label{sec:level2}Nonlinear Correlations (NLC) Scaled Strategy}
\begin{figure}[t!] 
  \centering
     \includegraphics[width=0.5\textwidth]{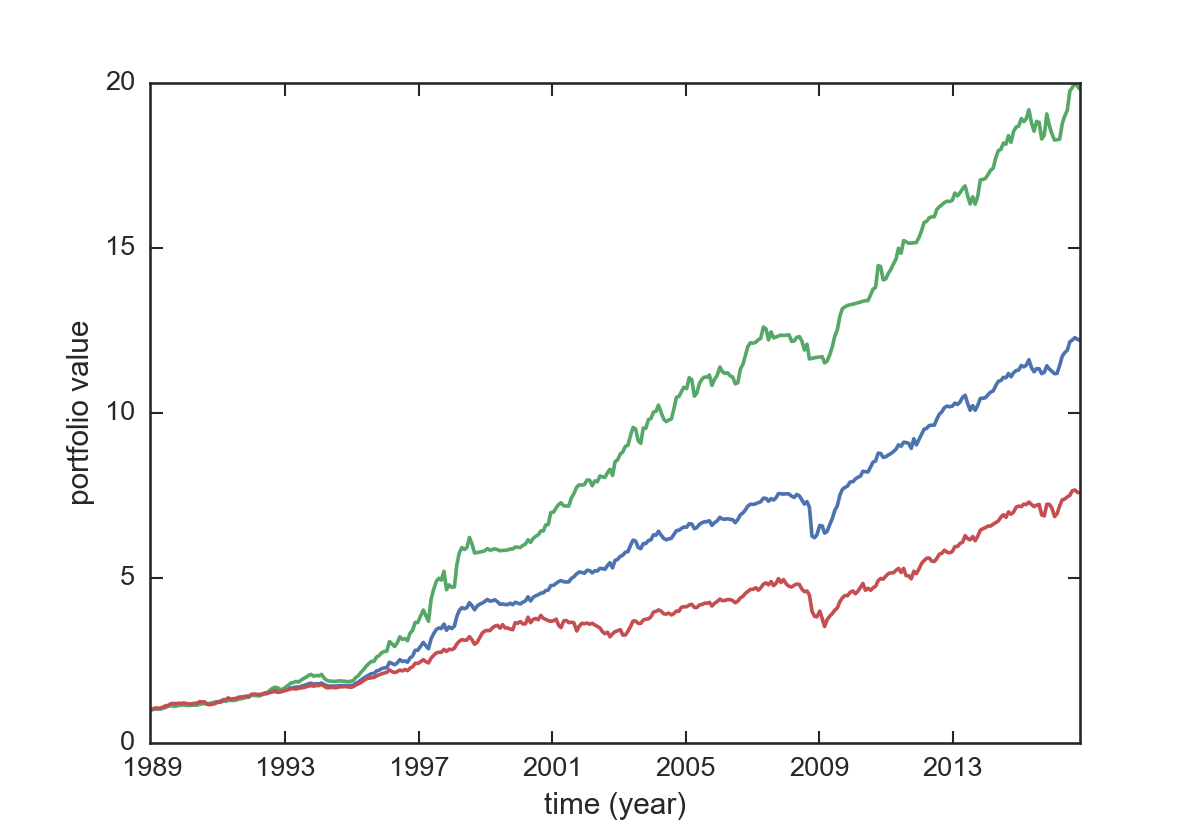}
  \caption{Backtest results: Portfolio value (starting at 1 at $t=0$) for fixed allocation (red), fully invested max sharp ratio portfolio (blue) and NLC scaled portfolio (green).}
  \label{fig:pfval}
\end{figure}
Having understood that there are significant nonlinear correlations present in stock market returns, naturally the question arises how we can make practical use of this knowledge. When it comes to portfolio optimization it is common standard to describe mutual interrelationships of assets through their Pearson correlation coefficients. As outlined above, a significant fraction of the information about mutual dependencies of stocks can be of nonlinear nature and hence is not captured by the linear Pearson coefficient. In times when nonlinear correlations are low, the linear correlation matrix captures most of the information and hence is a good estimator for interdependencies. When nonlinear correlations are high, however, a significant amount of information is missing and thus making it a bad estimator. The idea is now to perform a classical Markowitz optimization and chose the maximum sharp ratio portfolio as our benchmark portfolio. The alternative strategy takes the same relative allocation but allows for an additional asset: Cash. We allow cash weights from -100\% to 100\%. A weight of -100\% means we allow borrowing money in order to increase the investment exposure. For example, imagine we invest 1000\$ in our benchmark portfolio. A cash weight of -100\% in the alternative strategy then means that we borrow another 1000\$ in order to increase the investment exposure to 2000\$. In contrast, 100\% cash weight in the alternative strategy means that we only hold cash and do not invest into other assets i.e. have an investment exposure of 0\$. The cash weight is determined by a nonlinearity score $s_{nlc}$, which includes the following measures: \\
\\
1. Absolute strength of nonlinear correlations
\begin{eqnarray}
s_1(t)=  \langle \zeta_{nlc}(X_i,X_j, t) \rangle_{i,j} \ \ for \ i \neq j \ .
\label{eq:score1}
\end{eqnarray}
2. Nonlinear correlations versus a 24 time steps rolling window mean which corresponds to around two years
\begin{eqnarray}
s_2(t) = \frac{1}{24} \sum^{t}_{t^*=t-24}  s_1(t^*) \ .
\label{eq:score2}
\end{eqnarray} 
3. Change of nonlinear correlations versus a three time step rolling window corresponding to one quarter
\begin{eqnarray}
s_3(t) = \frac{s_1(t)}{s_1(t-3)} - 1 
\label{eq:score3}
\end{eqnarray}
The idea here is to not only use the strength of nonlinear correlations in the current time step but also reflect how they compare to a rolling mid-term average. In addition, we incorporate how the strength of nonlinear correlations changed within the last three months. The reason is that we want to capture the beginning of turbulent market periods as well as to achieve a market re-entry after turbulent periods are over and markets recover. 
\begin{figure}[t!] 
  \centering
     \includegraphics[width=0.5\textwidth]{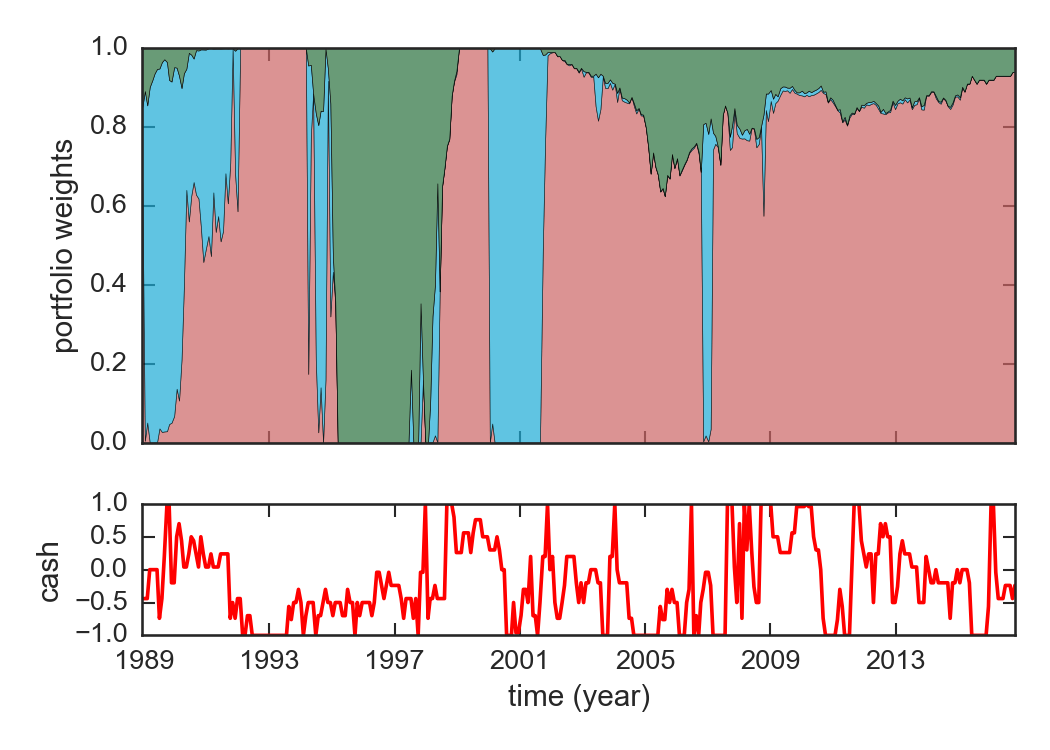}
  \caption{Upper plot: Portfolio weights --- S\&P 500 Index (green), Merrill Lynch US Corporate Index (red) and Merrill Lynch US Treasury Index (blue). Lower plot: Cash weights based on NLC strategy.}
  \label{fig:weights}
\end{figure}
\begin{table*}
\begin{ruledtabular}
\begin{tabular}{llllll}
&$0\leq s_1(t) < 0.1$& $0.1\leq s_1(t) < 0.15$&$0.15\leq s_1(t) < 0.2$ &$0.2\leq s_1(t) < 0.25$ &$0.25 \leq s_1(t)$ \\     
$s^*_1(t)$&$100$&$75$ &$50$ &$25$&$0$\\ \hline  \noalign{\vskip 0.05in}
&$0\leq s_1(t) < \frac{1}{2} s_2(t)$& $\frac{1}{2} s_2(t) \leq s_1(t) < \frac{3}{4} s_2(t)$&$\frac{3}{4} s_2(t)\leq s_1(t) < \frac{21}{20} s_2(t)$ &$\frac{21}{20} s_2(t)\leq s_1(t) < \frac{23}{20} s_2(t)$ &$\frac{23}{20} s_2(t) \leq s_1(t)$\\
$s^*_2(t)$&$100$&$75$ &$50$ &$25$&$0$\\ \hline  \noalign{\vskip 0.05in}
&$s_3(t) \leq -0.02$&$-0.02 \leq s_3(t) < 0$&$0 \leq s_3(t) < 0.02$  & $0.02 \leq s_3(t) < 0.05$&$0.05 \leq s_3(t) < 0$ \\
$s^*_3(t)$&$25$&$10$&$0$   & $-10$&$-100$ \\
\end{tabular}
\caption{Conversion of portfolio scoring measures $s_1(t), s_2(t), s_3(t)$ to scores $s^*_1(t), s^*_2(t), s^*_3(t)$ . }
\label{table:scoring}
\end{ruledtabular}
\end{table*}
The above measures are converted to scores as shown in Table~\ref{table:scoring} using a nonlinear mapping. The absolute strength of nonlinear correlations $s_1(t)$ translates to the most defensive score $s^*_1(t)=0$ if $s_1(t)\geq 0.25$. This means that on average 25\% or more of the mutual information of the assets is of nonlinear nature. Likewise, if the strength of nonlinear correlations $s_1(t)$ exceeds its two-year moving average $s_2(t)$ by 0.15 or more, the score $s^*_2(t)$ takes on its most defensive value $s^*_2(t)=0$. We built the scoring model such that a strong increase in $s_3(t)$ has a larger impact (-100) on the score $s^*_3(t)$ than a strong decrease (+25). This is because firmly growing nonlinear correlations might be a sign for turbulent market periods and in this case we would like to have high cash weights. All three measures are equally weighted and bounded between 0 and 100
\begin{eqnarray}
\tilde{s}_{nlc}(t) = \frac{1}{3}(s_1^*(t) + s_2^*(t) + s_3^*(t)) \\
s_{nlc}(t) = \max \{ \min \{ \tilde{s}_{nlc}(t), 100\},0\} \ .
\label{eq:weight}
\end{eqnarray}
The cash weight of portfolio is then determined by
\begin{eqnarray}
w_{cash}(t) =   \frac{1}{100}(100-2 s_{nlc}(t)) \ .
\label{eq:cashweight}
\end{eqnarray}
Hence, if nonlinear correlations are stronger, the linear correlation matrix captures less information about dependencies and thus the strategy leads to a more defensive allocation (higher cash exposure) and vice versa. The strategy likewise worked for different parameter choices but we decided to focus on the simple equally weighted method presented above. We use a 500 day sliding window for the calculation of $\zeta_{nlc}(X_i,X_j, t)$ and the covariance matrix for the Markowitz optimization in order to better capture changing market environments.

In order to test our strategy we selected a simple setup of three indices plus cash: Merrill Lynch US Corporate Index LOC (US Corporate Bond Index), Merrill Lynch US Treasury Index USD unhedged (US Government Bond Index), S\&P 500 Index (US Equity Index) and BBA LIBOR USD 1 Month (USD Cash Rate). The reason for not taking the large subset of stocks from the S\&P 500 Index we used in the first part of the study is that we achieve a more stable allocation over time in this simple example which is easier to interpret. When taking a large portfolio universe the allocations would tend to change a lot in each time step. Besides that we would like to have less risky assets than stocks in our portfolio universe such as US Government Bonds in order to see if our strategy achieves the right allocations during different market periods. We choose the indices above because they reflect a large amount of the US credit, government bonds and stock market.

We then ran a backtest from 1988 until 2016 performing a Markowitz optimization and portfolio rebalancing every 20 time steps. Figure~\ref{fig:pfval} shows the development of the fully invested zero cash portfolio (blue), a strategic asset allocation with constant 25\% Corporate Bonds, 25\% Government Bonds and 50\% Equities (SAA -- red) and finally our strategy with dynamic cash weights (green). In Fig. \ref{fig:weights} we see that our strategy leads to 100\% cash weight from September 2008 until February 2009 and thus achieved a safe allocation just before the crash commenced.
We notice that the strategy outperforms the zero cash strategy by 62\% and the fixed allocation even by 161\%. However, we have to account for the higher investment exposure of 120\% on average corresponding to minus 20\% cash leverage. For this we ran another backtest where we set a constant cash weight of minus 20\%. It turns out that our NLC strategy still outperforms by around 18\%. Thus we conclude that this outperformance is not driven by higher risk due to a higher investment exposure but the dynamics of the nonlinear correlations signal itself. 
\section{Summary and Conclusion}
\label{sec:summary}
In this study we analyzed nonlinear correlations in multidimensional financial time series by using mutual information as a measure for both linear and nonlinear dependencies and the method of surrogate data. In the first step we compared the moments of distance matrix coefficients obtained from mutual information to the results based on Pearson correlation. We found that especially during turbulent market periods e.g. the 2008 crisis both results show qualitative differences and hence indicate significant nonlinear correlations. This stands in contrast to the expectation that during crises dependencies reduce mainly to linear correlations. Then we constructed Minimum Spanning Trees and equivalently found differences in the network topology between the linear and nonlinear measure. It turned out that the average distance from the center of the network is significantly lower during periods of crises when considering nonlinear correlations. Furthermore, the center of the network in terms of degree centrality itself is more stable and less fluctuating in the linear case. We showed that the centrality of J.P.Morgan grew extensively long before the subprime mortgage bubble crashed in 2008. To investigate if such kind of measures could potentially act as an early warning indicator we will analyze the dynamics of the average centrality of industrial sectors in further studies. It is very interesting to understand that there are different types of financial crises in terms of nonlinear effects. The results of our study indicate that during the 2008 crisis nonlinear effects were significantly stronger than in preceding crises. Furthermore, we found that major political events seem to have no significant mid- and long-term impact on market correlation structure in contrast to financial crises. Finally, after understanding that there are significant nonlinear correlations present in stock returns we developed a practical application in the field of portfolio optimization. We showed that scaling the investment exposure based on the strength of nonlinear correlations leads to significant outperformance as compared to a fully invested portfolio. More direct applications of this knowledge will be explored in further studies.

\bibliography{NonlinearNetworks}

\end{document}